# A Scheme for Automation of Telecom Data Processing for Business Application

[1]T.R.Gopalakrishnan Nair, [2]Vithal. J. Sampagar, [3]Suma V, [4]Ezhilarasan Maharajan

[1, 3]Research and Industry Incubation Center, Dayananda Sagar Institutions, Bangalore, India
{Sumavdsce, trgnair }@gmail.com
[2]vithal.sampagar@gmail.com, Dayananda Sagar College of Engineering, Bangalore, India
[4]Razorsight Software Pvt Ltd, Koramangala Industrial area,
5th Block, Koramangala, Bangalore, India

**Abstract.** As the telecom industry is witnessing a large scale growth, one of the major challenges faced in the domain deals with the analysis and processing of telecom transactional data which are generated in large volumes by embedded system communication controllers having various functions. This paper deals with the analysis of such raw data files which are made up of the sequences of the tokens. It also depicts the method in which the files are parsed for extracting the information leading to the final storage in predefined data base tables. The parser is capable of reading the file in a line structured way and store the tokens into the predefined tables of data bases. The whole process is automated using the SSIS tools available in the SQL server. The log table is maintained in each step of the process which will enable tracking of the file for any risk mitigation. It can extract, transform and load data resulting in the processing.

## 1. Introduction

One of the significant operations of Telecom industry is the exchange or transfer of customer information between various modes of applications. Current industry scenario operates using file types such as Comma Separated Value (CSV), Exchange Message Interface (EMI) etc. for the analysis purpose. Having received the data, it will be processed and put into the database tables using various tools such as Extract Transform and Loading (ETL) tools, bulk insert query and tool, import functionality tool, which are available in various databases. However, Informatica, Data Stage and SQL Server Integration Service (SSIS) tool are the most commonly used ETL tools for data transformation and for the analysis purpose. Customer information is prepared in the line by line text format, which comprises of a sequence of tokens such as words. Tokens are extracted from the Comma Separated Value (CSV) file and stored into the database table. However, Parser is an application which is used to extract the information from the CSV files and store them into the database tables to perform different kinds of applications. The transformed data can be used for various applications such as analytical and data ware house applications. The analysis of customer information ensures the effectiveness of cost management and the revenue assurance of the industry.

CSV file has become the standard file format for the Telecom Service providers to exchange the data of different types and of different formats. The file may contain customer information of the telecom Service provider. However, there are certain industry standard rules for processing these files. According to [1] [2], Comma Delimited file or .CSV file is a data set where every record is separated by comma and every new line represents a new row / record in the database table. The data or token are values which are organized into cells or fields via comma separators. The first lines of the file are the optional column header field. Absence of header indicates the first data record or row of the file. According to the industry standard rule, blank field indicates absence of data, which is enclosed using double quotes (" "). The parser does not populate a blank field entry within double quotes into the database table [1], [2].

.CSV files are most frequently used to export database information which includes the log of database information. CSV files are one of the most appropriate file formats used by many telecom service providers to transmit information due to its low bandwidth consumption [5]. Authors in [8] suggest the use of CSV files in the decommutating process where the data is are stored in CSV (comma separated values) format files. They further recommend the users to configure and use other analysis tools to do further analysis of the data. However, author in [10] suggest the implementation of parser for pipe delimited file which has the file extension as .CSV. The author provides explanation on CSV file reading techniques using tools. Fig 1. depicts a sample CSV data.

```
"aaa","b <CR><LF>
bb","ccc" <CR><LF>
zzz, yyy,xxx
```

Fig.1 Sample CSV data

CSV files are popularly known as "Flat Files", since flat files contain a single table with finite number of rows and columns. Any application which uses a table to store its information and when the data base table is exported with an extension as .CSV, the file will be "flattened" with comma as a delimiter. Since, the data is stored in the text format, it can be opened in the any text formats such as Notepad, WordPad, Text pad, Microsoft excel etc. Authors in [5] express that currently .CSV files are commonly used on all the platforms in the telecom industry. They further state that the data can be transferred in program compatibility format [5]. Hence, the use of the CSV file minimizes the bandwidth usage and reduces the network traffic. The simple CSV file is illustrated in the Fig.2.

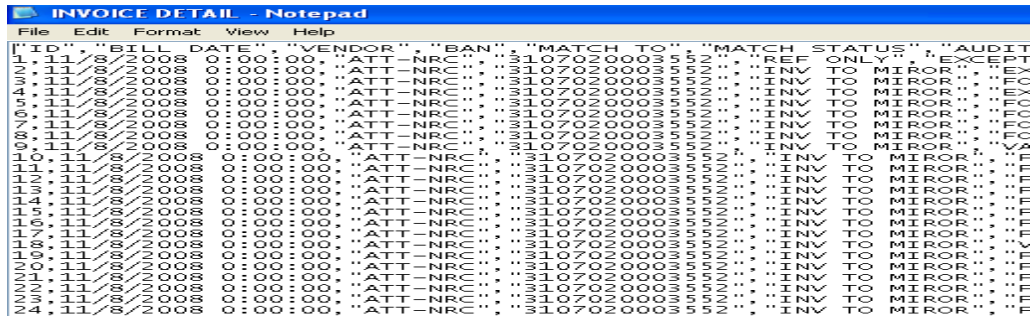

Fig.2 Sample CSV file

The CSV file format is simple and is supported by most of the spreadsheets and database management systems. Many programming languages have libraries available that support CSV files. Even modern software applications support CSV imports and/or exports as the format is widely recognized. Additionally, many applications allow .CSV-named files to use any delimiter character.

Nevertheless, the existing built in functions in the database such as Import function of the SQL Server and Bulk Insert of SQL Server, Oracle database are capable of extracting the information from the simple CSV files. However, these functionalities cannot process the complex CSV files which contains embedded comma and embedded line breaks. Most of the files which are currently being used in the telecom industry are complex CSV files. Therefore, there is a strong need to parser the complex CSV file using automation mode without manual intervene in order to reduce time and resource consumption. It increases availability of the parser and thereby increases productivity, quality and efficiency of the organization.

This paper introduces the automation of CSV file parser which can read the data from the CSV files and provide the interface to the user application to access the files. CSV parser is the software modules which are capable of handling these tasks. Due to the probability of processing the duplicate file with repeated information more than once, the automated parser further checks for the file states. It thus, identifies for the existence of duplicate file and indicates the same in the log table. This parser checks for the presence of the header in file since header describes the fields in the file and some files may or may not contain the header. The parser will parse the entire file in the absence of header. However, it ignores the header line in their presence and continues to parse the rest of the information. In either of the aforementioned conditions, the automated parser reads the CSV file and stores the processed result into the database. The parser will maintain the log table throughout the process in order to track the file in case of the failure.

In this paper, section 2 explains the existing methods of file processing in telecom industry. Section 3 is the designed methodology which briefs about the design followed for the implementation of automated parser along with the assumptions made in this work. Section 4 explains the automated CSV parser. Section 5 provides description about the integration environments which is required for the implementation of automated parser. Section 6 illustrates the results and section 7 provides the summary of the entire work.

## 2. Existing File Processing Techniques

Author in [7] explains the import task available in the Microsoft SQL server 2005 version onwards. He further provides details about the use of import functionalities to process simple CSV files into the database. The application will split the data based on the Delimiter (i.e. Comma). The import option available in the Microsoft SQL server as explained by the author in [7] however does not process the complex CSV files [7]. However, author in [9] explains parser application in Microsoft excel. He elucidates the process of extraction of information from the spreadsheets such as the Microsoft excel and further suggest to automate the process. The screen shot of the Import task in SQL server database is shown in the Fig.3.

The complex CSV files are those which contain the embedded commas, embedded space, embedded double quotes etc. The import process doesn't take care of the exceptions which may occur in the files. The import function splits the line based on the comma delimiter and does not consider the embedded comma. However, in the present situation, the industry uses very complex CSV files. The import functionality therefore fails to process such complex files. Additionally, the import function processes a single file at a time. Subsequently, it requires the dedicated resource to input the file and monitor the overall process.

BULK INSERT statement available in the SQL Server as well as the Oracle is used to process the files based on the Delimiter character [1] [2]. The Bulk insert statement of the SQL Server 2008 R2 is shown below

BULK INSERT database

FROM 'file location'

WITH

{

    FIELDTERMINATOR='delimiter character'

    ROWTERMINATOR='row terminator'

}

The use of aforementioned statement is feasible if it is simple CSV file and does not contain double quotes or embedded comma. This is because, BULK INSERT statement does not consider the embedded comma and other special characters that the user may include along with the various records such as "$" in the field. With the failure of processing single record by the BULK INSERT statement, the entire process is also failed. Yet another drawback of BULK INSERT statement is manual configuration of the file name and the path in the statement by the user.

Thus, the major drawback of existing technique of import function is that it considers a single file as an input. Multiple file processing is a tedious job with the bulk insert as the user is compelled to manually reconfigure the file path for every single file. The above specified major drawbacks motivated the implementation of automated parser for complex CSV files.

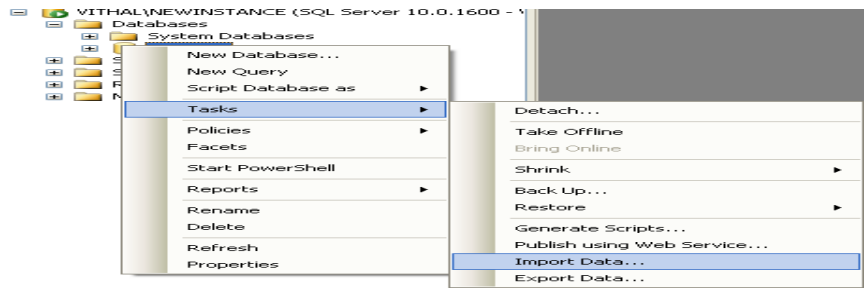

Fig.3 Import task in SQL Server Database.

1. **Design Methodology**

The work is carried out in a leading SAS 7 certified telecom industry. The company provides analytical solutions to the major leading North American Telecom service providers. The company operates on Microsoft products and on object technology. The company used to function using import function of SSIS tool with dedicated human resource to parse the CSV files. This mode of operations was expensive in terms of time and resources. This investigation was carried out in order to reduce expensive nature of production and increase productivity with reduced time, cost and resources. The automated parser operates only on Microsoft technologies such as SQL server, SSIS, C # and .Net language. The work was carried out using the software development life cycle. Requirements were collected from various telecom industry vendors and were appropriately analyzed. Certain assumptions were made during the design and development of the system such as listed below:

- Assumption 1- file does not contain line breaks
- Assumption 2 - column header does not contain any numeric data and is purely a string.
- Assumption 3 - The records are in the form of rows and columns
- Assumption 4 - The generic table which is used to store the parsing results is previously developed in the database. This table will hold the parsing result of the all the files.

High level system design was developed and was implemented using C # and .Net. The entire development code is integrated to SSIS tool for the automation purpose. Several tests such as unit test, integration test, system test, regression test were performed. The developed automated parser was installed in the ETL environment.

## 2. Automated CSV Parser

This work is the automation of the Telecom File Processing that can parse the file having millions of records without manual intervenes. The details of the each file are stored into the log table.

The automated CSV parser can parse all types of CSV files. Prior to the parsing, the process such as file status check, header check and duplicate record check are performed. A detail of every file which is parsed by the parser is maintained in the log table. The log table thus enables one to track the file location during the process failure. Automated CSV parser creates folders automatically (In, In Progress, Archive, and Exception) in the absence of availability of such folders in the C drive. Files which are required to be processed are stored into the Source folder which acts as an input for the project. A log is maintained at every stage of parsing. The automated parser consists of three stages namely header check stage, duplicate file check and parsing stage.

The first stage of automated parsing is the header check stage which checks for the header status. In this stage, the first line from the file is read and checked for the presence of alphanumeric string. The header describes the fields that are present in the file. It is assumed that the header is purely string and does not contain any numeric value. The presence of numeric value is not treated as a header in the file. However, the log table is updated for the file name which is already recorded into the database table. The file location is also maintained during each step. All the files will be moved to the "In" folder after the completion of this step. And the file location is updated.

Subsequent task is the second stage of automated parsing which checks for the duplicate record in the file. With the arrival of a new file from the In folder, first 10 rows of the file is read into a table called as reference table in the database. Each time a file is about to process, the extracted values from the file are compared with the data present in the reference table. If the comparison yields common value, it indicates that the file is previously processed and hence ignores the file. The file is henceforth moved to the exception folder. The new location of the file is updated at the log table and the file is tagged as a duplicate. Another advantage of this process is to find error status of the file. One of the filed of the file contains critical values which are critical and most significant information of the record. Null value in the critical column indicates that the file is and is thus moved to the exception file folder. The files which are not considered as duplicate files are moved to the "In Progress" folder, which acts as an input to the parser module.

The last stage of automated parser is to parse the files. At this stage, StreamReadear, an inbuilt function is used to read a file line by line. SreamReader increases the performance of the parser. Each line is parsed using the parser logic and stored into the array object. The array object is created from the parsed data which contains the entire token list of a single line that has to be stored into database. The array object is bounded with some threshold value. The database connectivity is made and the entire data is pushed to the database table. Once the array reaches that threshold, the entire content of the array is moved to the database table. The array object is created by using the count of the header elements. When the parser encounters a quotation mark, the entire token until the next quotation mark is read and the extracted tokens are stored in the array objects. This process is repeated until the end of the file is encountered. Lastly, the number of the columns in the file as well as the number of the rows processed is updated in the log file. The automated parser sends the email notification to the concerned person in case of both the process failure and with the encounter of duplicate file. An email notification is also sent with successful completion of the parsing process. Fig. 4. depicts the data flow chart of the automated parser.

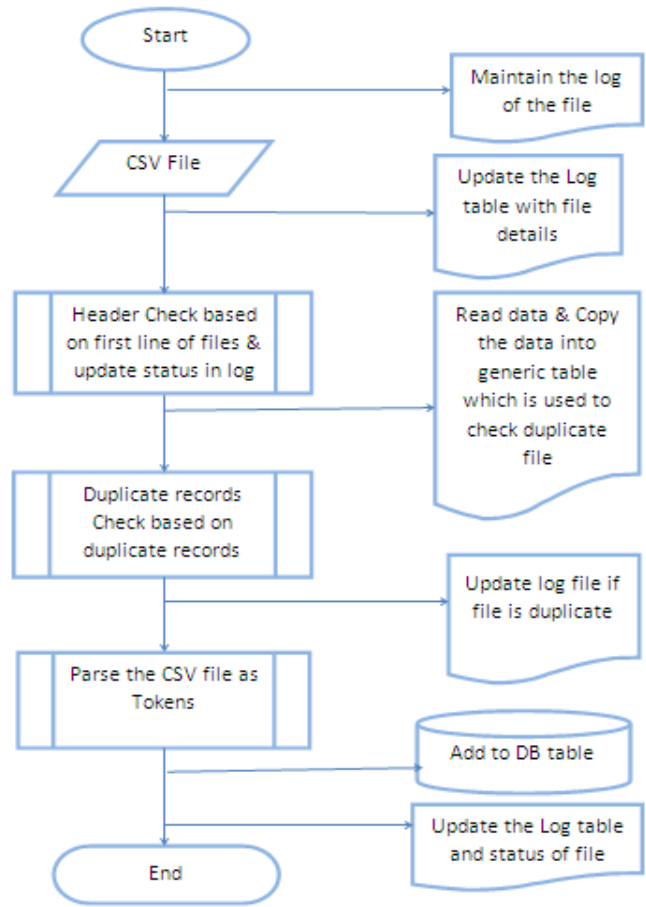

Fig.4 Parser Flowchart

### 3. Integration Environment

Author in [6] provides explanation on the flexibility of the SSIS and the challenges that are faced in business. He suggests that the data integration technologies provide meaningful, reliable information to maintain a competitive advantage in today's business world [6].

Major data source in the telecom domain includes the information in the form of CSV files which transmits the data for different applications between different locations. The parser is integrated to the SQL Server Integration service (SSIS) environment in order to automate the entire parsing process. SSIS tool is a platform for data integration and workflow applications. It contains features which can perform data warehousing applications in a faster and flexible mode of data extraction, transformation, and for loading (ETL) purpose. The tool further can be used to automate maintenance of SQL Server databases and for updates to multidimensional cube data. SSIS enables data migration task. The automated parser is thus the standard tool in the any data ware house related applications.

### 4. Results

The CSV parser introduced in this paper is implemented using C# and .NET in SharpDevelop 4.0 open Source IDE. The entire parser is integrated to the SQL Server Integration Service (SSIS) to automate the process. This section illustrates the result of the automated CSV Parser. The input file is shown in the Fig.5. The CSV files are stored in the source folder which contains all the files required for processing. The folder may contain as many files as external application requires. The result of the parsing is shown in the Fig.6.

The result of the parsing can be used for the analysis and the data ware house applications. The parser represented in this paper is limited to the file size of 250MB.

The CSV parser introduced in this paper can be enhanced further for the other type of files such as Tab Delimited Files as well as Text files.

Fig.5 Input CSV files

The screen shot of the database table is shown in the Fig.6. The file contains header and hence is ignored while processing which is as observed in the database table. The table contains values from the second row onwards. Each row of the table indicates a line from the CSV file. The entire line is parsed and inserted into the database table. A record with no value is left blank during the insertion.

Fig.6 Database table with parsed values

The parser maintains a log of each file which is processed. The log table also contains number count of column and rows which are present in the flat file and which are processed into the database tables. Further, the log table provides information of the columns that indicates Header, file name, file path, file status, created and modified dates. These columns are dynamically updated as the file gets processed. The parser has the unique capability of parsing millions of records and stores into the reference table. The log table is show in the Fig.7

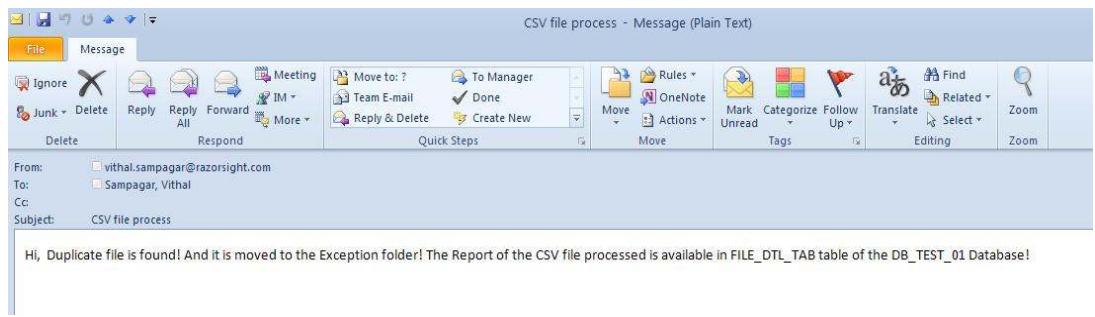

Fig.7 Log table

### 5.1 Email Notification

The automated parser is designed to send the email notification whenever there is an occurrence of any type of error such as existence of duplicate file during processing and for all successful completion of the parsing process. The email notification is shown in Fig.8.

Fig.8 Email Notification

## 5. Conclusion

The parser is an application for the SSIS tool, which is used for analyzing the telecom file consisting of sequence of the tokens. It extracts the information from the file and stores the data into predefined database. Comma Delimited files (CSV) file has become the standard file format for the Telecom Service providers to exchange the data in different types and of different formats. However, parser existing in the current telecom industry is manual driven and is expensive in terms of time, resource and cost. Further, industry operates on complex CSV files which contain several primitives and the currently available parsers are only capable of parsing simple CSV files.

This application is useful in the Extract, Transform and Loading (ETL) process of data ware house and the business intelligence applications. It is integrated to the SSIS tool of ETL process. The application is tested with various valid CSV files and the results strongly prove that it can be used in ETL process. However, the automated parser operates only on Microsoft technologies. It can handle all the exceptions where as the import function available in the SQL Server, the Bulk insert of the SQL Server as well as Oracle cannot handle such things. The stored data is here analyzed for different applications such as calculation of overall revenue, cost management etc. The extracted values are also used for various data ware house applications. Thus, automated parser is shown to be parsing complex CSV file using the auto mode without manual intervention in order to reduce consumption of time and resource. It increases availability of this efficient tool and thereby increases productivity, quality and efficacy of the organization.

## Acknowledgement

Author would like to thank Razorsight Software Pvt. Ltd and the Telecom domain industry experts for providing an opportunity to carry out this part of the research work successfully.